\documentclass[hyper]{JHEP} % 10pt is ignored!

\usepackage{epsfig}
\newcommand\fverb{\setbox\pippobox=\hbox\bgroup\verb}

\newcommand\fverbdo{\egroup\medskip\noindent%

            \fbox{\unhbox\pippobox}\ }

\newcommand\fverbit{\egroup\item[\fbox{\unhbox\pippobox}]}

\newbox\pippobox

\title{Comments About Hamiltonian Formulation of
 Non-Linear Massive Gravity with
St\"{u}ckelberg Fields }
\author{J. Kluso\v{n}\\
Department of
Theoretical Physics and Astrophysics\\
Faculty of Science, Masaryk University\\
Kotl\'{a}\v{r}sk\'{a} 2, 611 37, Brno\\
Czech Republic\\
E-mail: \email{klu@physics.muni.cz}}
\preprint{}

 \abstract{We perform the Hamiltonian analysis of
 some form of the
 non-linear massive gravity action that is   formulated in
 the St\"{u}ckelberg formalism.
Following seminal analysis performed in arXiv:1203.5283 [hep-th] we
find
 that this theory possesses one primary
 constraint which could eliminate one
 additional mode in this theory. We
 performed the explicit Hamiltonian
 analysis of two dimensional non-linear
 massive gravity and we found that this
is theory free from the ghosts. }

\def\bA{\mathbf{A}}
\def\bB{\mathbf{B}}

\def\mC{\mathcal{C}}
\def\be{\begin{equation}}

\def\ee{\end{equation}}

\def\bea{\begin{eqnarray}}

\def\eea{\end{eqnarray}}

\def\tr{\mathrm{tr}\, }

\def\mH{\mathcal{H}}

\def\bz{\mathbf{z}}

\def\tr{\mathrm{Tr}}

\def\bx{\mathbf{x}}
\def\by{\mathbf{y}}

\newcommand{\hg}{\hat{g}}

\newcommand{\mK}{\mathcal{K}}
\newcommand{\hmK}{\hat{\mK}}

\newcommand{\mU}{\mathcal{U}}

\newcommand{\mG}{\mathcal{G}}

\def \bA{\mathbf{A}}

\newcommand{\bT}{\mathbf{T}}

\newcommand{\mL}{\mathcal{L}}

\def\pb #1{\left\{#1\right\}}

\begin{document}
%%%%%%%%%%%%%%%%%%%%%
%%%%Introduction %%%%%%%%%
%%%%%%%%%%%%%%%%%%%%
\section{Introduction}\label{first}
One of the most challenging problem is
to find consistent formulation of
massive gravity. The first attempt for
construction of  this theory is dated
to the year 1939 when Fierz and Pauli
formulated its version of linear
massive gravity \cite{Fierz:1939ix}
\footnote{For review, see
\cite{Hinterbichler:2011tt}.}. However
it is very non-trivial task to find  a
consistent non-linear generalization of
given theory and it remains as an
intriguing theoretical problem. It is
also important to stress that recent
discovery of dark energy and associated
cosmological constant problem has
prompted investigations in the long
distance modifications of General
Relativity , for review, see
\cite{Clifton:2011jh}.

Returning to the theories of massive
gravity we should mention that these
theories suffer from the problem of the
ghost instability, for very nice
review, see \cite{Rubakov:2008nh}.
Since the General Relativity  is
completely constrained system there are
four constraint equations along the
four general coordinate transformations
that enable to eliminate four of the
six propagating modes of the metric,
where the propagating mode corresponds
to a pair of conjugate variables.  As a
result the number of physical degrees
of freedom is equal to two which
corresponds to the massless graviton
degrees of freedom. On the other hand
in case of the massive gravity the
diffeomorphism invariance is lost and
hence the theory contains six
propagating degrees of freedom which
only five correspond to the physical
polarizations of the massive graviton
while the additional mode is ghost.

It is natural to  ask the question
whether it is possible to construct
theory of massive gravity where one of
the constraint equation and associated
secondary constraint eliminates the
propagating scalar mode. It is
remarkable that linear Fierz-Pauli
theory does not suffer from the
presence of such a ghost. On the other
hand it was shown by Boulware and Deser
\cite{Boulware:1973my} that ghosts
generically reappear at  the non-linear
level. On the other hand it was shown
recently by de Rham and Gobadadze in
\cite{deRham:2010ik} that it is
possible to find such a formulation of
the massive gravity which is ghost free
in the decoupling limit. Then it was
shown in \cite{deRham:2010kj} that this
action that was written in the
perturbative form can be resumed into
fully non-linear actions \footnote{For
related works, see
\cite{Mirbabayi:2011xg,Sjors:2011iv,Burrage:2011cr,Gumrukcuoglu:2011zh,Berezhiani:2011mt,
Comelli:2011zm,Kluson:2011aq,Comelli:2011wq,Mohseni:2011vv,Gumrukcuoglu:2011ew,Hassan:2011zd,
Hassan:2011tf,D'Amico:2011jj,deRham:2011qq,deRham:2011pt,Gruzinov:2011mm,Koyama:2011yg,
Nieuwenhuizen:2011sq,
Koyama:2011xz,Chamseddine:2011bu,Volkov:2011an}.}.
The general analysis of the constraints
of given theory has been performed
 in \cite{Hassan:2011hr}. It was argued
 there that it is possible to perform
 such a redefinition of the shift
 function so that the
resulting theory still contains the
Hamiltonian constraint. Then it was
argued that the presence of this
constraint allows to eliminate the
scalar mode and hence the resulting
theory is the ghost free massive
gravity. However this analysis was
questioned in \cite{Kluson:2011qe}
where it was argued that it is possible
that this constraint is the second
class constraint so that the phase
space of given theory would be odd
dimensional. On the other hand in the
recent paper \cite{Hassan:2011ea} very
nice analysis of the Hamiltonian
formulation of the most general gauge
fixed non-linear massive gravity
actions was performed  with an
important conclusion that the
Hamiltonian constraints has zero
Poisson brackets. Then the requirement
of the preservation of this constraint
during the time evolution of the system
implies an additional constraint. As a
result given theory has the right
number of constraints for the
construction of non-linear massive
gravity without additional scalar mode
\footnote{Alternative arguments for the
existence of an additional constraints
were given in \cite{Golovnev:2011nz}
even if the Hamiltonian analysis was
not complete and the minimal non-linear
massive gravity action was considered
only.}.

All these results suggest that the gauge fixed form of the
non-linear massive gravity actions could be ghost free theory. On
the other hand the manifest diffeomorphism invariance is lacking and
one would like to confirm the same result in the gauge invariant
formulation of the massive gravity action using the St\"{u}ckelberg
fields. In fact, it was argued in \cite{deRham:2011rn} that for some
special cases  such a theory possesses an additional primary
constraint whose presence implies such a constraint structure of
given theory that could eliminate one additional scalar mode.
However the limitation of this analysis is that it was not performed
for the general metric so that one can still ask the question
whether the elimination of the ghost mode occurs in general case as
well. In fact, there is a well known example of the theory which
seems to be consistent around some background while it is
pathological for general background which is the Fierz-Paull theory
\cite{Fierz:1939ix} which is ghost free around the flat background
while it contains ghosts in general background
\cite{Boulware:1973my}. This paper is the first step in the
Hamiltonian analysis of St\"{u}ckelberg formulation of the
non-linear massive gravity with general metric. More precisely, we
consider one particular class of the non-linear massive action that
was presented in \cite{Hassan:2011hr,Hassan:2011vm}. This action is
sufficiently simple to be able to find the explicit relation between
time derivatives of scalar fields and conjugate momenta while it
possesses all interesting properties of the non-linear massive
gravity as was shown in \cite{Hassan:2011ea,Golovnev:2011nz}
\footnote{We should stress that the bimetric non-linear massive
gravity is also manifestly covariant form of the non-linear massive
gravity action \cite{Hassan:2011zd,Hassan:2011tf,Hassan:2011ea}. In
fact, it was argued in these papers that these manifestly
diffeomorphism invariant massive  gravities are ghost free as
well.}.

We find the Hamiltonian form of given action and determine primary
and secondary constraints of given theory. Then we show that due to
the structure of the non-linear massive gravity action this theory
possesses one primary constraint that is a consequence of the fact
that $\det V^{AB}=\det (\partial_i \phi^A\partial_j \phi^B
g^{ij})=0$) as was firstly shown in \cite{Hassan:2012qv}
\footnote{We should stress that in the previous versions of given
paper the constraint $\det V^{AB}=0$ was not taken into account and
hence wrong conclusions were reached.} . This result has a crucial
consequence for the structure of the theory. On the one hand this
constraint could provide a mechanism for the elimination of an
additional scalar mode however on the other hand the condition $\det
V^{AB}=0$ makes the calculation of the algebra of the Hamiltonian
constraints very difficult. In fact, we were not able to calculate
this algebra in the full generality with exception of the two
dimensional case where however two dimensional gravity is trivial.
More precisely, the Hamiltonian treatment of two dimensional
non-linear massive gravity shows that there are no physical degrees
of freedom left and this result coincides with the analysis
performed recently in \cite{deRham:2011rn,Kluson:2011aq}.

The structure of this paper is as
follows. In the next section
(\ref{second}) we review the main
properties of non-linear massive
gravity and rewrite it into more
tractable form that is suitable for the
Hamiltonian analysis which will be
performed in section (\ref{third}). In
section (\ref{fifth}) we perform the
Hamiltonian analysis of two dimensional
non-linear massive gravity action. In
conclusion (\ref{sixth}) we outline our
results and suggest possible extension
of this work. Finally in the Appendix
(\ref{appendix}) we perform the
calculation of the Poisson brackets of
the constraints when we do not impose
the condition $\det V^{AB}=0$. Of
course this is not the case of the
non-linear massive gravity action but
we include this appendix in order to
show the complexity of given
calculation even if we should again
stress that this should be considered
as a toy model calculation.

%%%%%%%%%%%%%%%%%%%%%%%%%%%%%%%%%%%%%%%%%%%%%%
\section{Non-Linear Massive Gravity}\label{second}
Our goal is to study non-linear massive
gravity action \cite{deRham:2010kj} in
the version that appears in
\cite{Hassan:2011hr,Hassan:2011vm}
\begin{equation}\label{Smassive}
S=M_p^2\int d^4x
\sqrt{-\hg}{}^{(4)}R(\hg)-
\frac{1}{4}M_p^2 m^2\int d^4x
\sqrt{-\hg}\mU (\hg^{-1}f) \ .
\end{equation}
Note that by definition $\hg^{\mu\nu}$
and $f_{\mu\nu}$ transform under
general diffeomorphism transformations
$x'^\mu=x'^\mu(x)$ as
\begin{equation}
\hg'^{\mu\nu}(x')= \hg^{\rho\sigma}(x)
\frac{\partial x'^\mu}{\partial x^\rho}
\frac{\partial x'^\nu}{\partial
x^\sigma} \ , \quad f'_{\mu\nu}(x')=
f_{\rho\sigma}(x)\frac{\partial
x^\rho}{\partial x'^\mu}\frac{\partial
x^\sigma}{\partial x'^\nu} \  .
\end{equation}
Now the requirement that the
combination $\hg^{-1} f$ has to be
diffeomorphism invariant implies that
the potential $\mU$ has to contain the
trace over space-time indices. Further,
it is convenient to parameterize the
tensor $f_{\mu\nu}$ using four scalar
fields $\phi^A$ and some fixed
auxiliary metric
$\bar{f}_{\mu\nu}(\phi)$ so that
\begin{equation}
f_{\mu\nu}=\partial_\mu\phi^A\partial_\nu
\phi^B \bar{f}_{AB}(\phi) \ ,
\end{equation}
where the metric $f_{AB}$ is invariant
under diffeomorphism transformation
$x'^\mu=x^\mu(x')$ which however
transforms as a tensor under
reparametrizations of $\phi^A$. In what
follows we consider
$\bar{f}_{AB}=\eta_{AB}$, where
$\eta_{AB}=\mathrm{diag}(-1,1,1,1)$.

The fundamental ingredient of the
non-linear massive gravity is the
potential term. The most general forms
of this potential were derived in
\cite{Hassan:2011vm,deRham:2010kj}. Let
us consider the minimal form of the
potential introduced in
\cite{deRham:2010kj}
\begin{eqnarray}
\mathcal{U}(g,H)&=&
-4\left(\left<\mK\right>^2-\left<\mK^2\right>\right)
=\nonumber \\
&=&
 -4\left( \sum_{n\geq
1} d_n \left<H^n\right>\right)^2-
8\sum_{n\geq 2} d_n\left<H^n\right> \ ,
\nonumber \\
\end{eqnarray}
where we now have
\begin{eqnarray}
H_{\mu\nu}&=&\hg_{\mu\nu}-
\partial_\mu\phi^A\partial_\nu\phi^B\eta_{AB}
\ , \quad
H^\mu_\nu=\hg^{\mu\alpha}H_{\alpha\nu}
\ , \nonumber \\
\mK^\mu_\nu&=&\delta^\mu_\nu-\sqrt{\delta^\mu_\nu-
H^\mu_\nu}=-\sum_{n=1}^\infty d_n
(H^n)^\mu_\nu \ , \quad
d_n=\frac{(2n)!} {(1-2n)(n!)^24^n} \ .
\nonumber \\
\end{eqnarray}
and where $(H^n)^\mu_\nu=
H^\mu_{\alpha_1}H^{\alpha_1}_{\alpha_2}
\dots H^{\alpha_{n-1}}_\nu$. Using this
explicit form of the potential we
present an important observation
related to the fact that the potential
has to be defined using the trace over
curved space-time indices which is a
consequence of the requirement of the
diffeomorphism invariance of the
non-linear massive gravity action
written in the St\"{u}ckelberg field.
Then we observe
\begin{eqnarray}
H^\mu_\mu &=& \delta^\mu_\mu-
\hg^{\mu\rho}\partial_\rho \phi^A
\partial_\mu \phi_A\equiv
\delta^A_{ \ A}-\bA^A_{ \ A}= \mH^A_{ \
A} \ , \quad \mH^A_{\ B}=\delta^A_{ \
B}-\hg^{\mu\nu}\partial_\mu
\phi^A\partial_\nu\phi_B \ ,
 \nonumber \\
 H^\mu_\nu H^\nu_\mu&=&
 \delta^\mu_\mu-\eta^{\mu\nu}
 \partial_\mu\phi^A\partial_\nu\phi_A-
 \eta^{\mu\nu}\partial_\nu\phi^A\partial_\mu\phi_A+
 (\eta^{\mu\nu}\partial_\nu\phi^A\partial_\mu\phi_B)
 (\eta^{\sigma\nu}\partial_\sigma\phi^B\partial_\nu\phi_A)=
\nonumber \\
&=&(\delta^A_{ \ B}-\bA^A_{ \ B})
(\delta^B_{ \ A}-\bA^B_{ \ A})=\mH^A_{
\ B}\mH^B_{ \ A} \ ,
\nonumber \\
 \left<H^n\right>^\mu_\mu&=&
H^{\mu}_{\nu_1} H^{\nu_1}_{\mu_2}\dots
H^{\mu_{n-1}}_{\nu}=
\mH^A_{B_1}\mH^{B_1}_{B_2} \dots
\mH^{B_{n-1}}_{ \ A} \ .
 \nonumber \\
\end{eqnarray}
These observation implies that the
potential can be written in the form
\begin{eqnarray}
\mU=-4\left(\sum_{n\geq 1}d_n
<\mH^n>\right)^2- 8\sum_{n\geq 2}
d_n<\mH^n>= \nonumber \\
=-4(<\hmK>^2-<\hmK^2>)= -4(\hmK^A_{ \
A})^2+4\hmK^A_{\ B}\hmK^B_{ \ A} \ ,
 \nonumber \\
\end{eqnarray}
where we defined
\begin{equation}
\hmK^A_{ \ B}=\delta^A_{ \
B}-\sqrt{\delta^A_{ \ B}-\bA^A_{ \ B}}
 \ .
\end{equation}
We mean that the trace over Lorentz
indices is more convenient for the
Hamiltonian treatment since we can
easily implement the ADM decomposition
of the space time metric.
 Explicitly,
we use  $3+1$ notation
\cite{Arnowitt:1962hi} \footnote{For
review, see \cite{Gourgoulhon:2007ue}.}
and write the four dimensional metric
components as
\begin{eqnarray}
\hat{g}_{00}=-N^2+N_i g^{ij}N_j \ ,
\quad \hat{g}_{0i}=N_i \ , \quad
\hat{g}_{ij}=g_{ij} \ ,
\nonumber \\
\hat{g}^{00}=-\frac{1}{N^2} \ , \quad
\hat{g}^{0i}=\frac{N^i}{N^2} \ , \quad
\hat{g}^{ij}=g^{ij}-\frac{N^i N^j}{N^2}
\ .
\nonumber \\
\end{eqnarray}
Note also that $4-$dimensional scalar
curvature has following decomposition
\begin{equation}\label{Rdecom}
{}^{(4)}R=K_{ij}\mG^{ijkl}K_{kl}+{}^{(3)}R
\ ,
\end{equation}
where ${}^{(3)}R$ is three-dimensional
spatial curvature, $K_{ij}$ is
extrinsic curvature defined as
\begin{equation}
K_{ij}=\frac{1}{2N} (\partial_t g_{ij}
-\nabla_i N_j-\nabla_j N_i) \ ,
\end{equation}
where $\nabla_i$ is covariant
derivative built from the metric
components $g_{ij}$. Note also that
$\mG^{ijkl}$ is de Witt metric defined
as
\begin{equation}
\mG^{ijkl}=\frac{1}{2}(g^{ik}g^{jl}+g^{il}g^{jk})-g^{ij}
g^{kl} \ , \quad
\mG_{ijkl}=\frac{1}{2}
(g_{ik}g_{jl}+g_{il}g_{jk})-\frac{1}{2}g_{ij}g_{kl}
\ .
\end{equation}
Finally note that in (\ref{Rdecom}) we
omitted terms proportional to the
covariant derivatives which induce the
boundary terms that vanish for suitable
chosen boundary conditions. Using this
notation we have
\begin{equation}
\bA^A_{ \ B}=
-\nabla_n\phi^A\nabla_n\phi_B+g^{ij}\partial_i\phi^A
\partial_j\phi_B \ ,
\quad \nabla_n\phi^A= \frac{1}{N}
(\partial_t\phi^A-N^i\partial_i \phi^A)
\ .
\end{equation}
We would like to stress that there is
an another issue with the construction
of the Hamiltonian formalism. To see
this note that the general potential
term contains the matrix $\sqrt{\bA}$
that can be defined as power series
$\sqrt{\bA^A_{ \ B}} =\sum_n c_n
(\bA^n)^A_{ \ B}$ with appropriate coefficients
$c_n$. Then the variation
of this expression is equal to
\begin{equation}
\delta \sqrt{\bA^A_{ \ B}}=c_1 \delta
\bA^A_{ \ B}+c_1 (\delta \bA^A_{ \ C}
\bA^C_{ \ B}+\bA^A_{ \ C}\delta \bA^C_{
\ B})+\dots
\end{equation}
that clearly cannot be written as $\delta \sqrt{\bA^A_{ \
B}}=\frac{1}{2} \delta \bA^A_{ \
C}\left((\sqrt{\bA})^{-1}\right)^C_{ \ B}$. However the situation
improves when we consider the potential term that depends on
$\sqrt{\bA^A_{ \ A}}$ since then we have \footnote{Note that this
result is consistent with the definition of the square root of the
matrix given in \cite{deRham:2010kj,Hassan:2011vm}
\begin{equation}
(\sqrt{\bA})^A_{\ B}(\sqrt{\bA})^B_{ \
C}= \bA^A_{ \ C} \
\end{equation}
since
\begin{equation}
\delta (\sqrt{\bA})^A_{ \ B}+
(\sqrt{\bA})^A_{ \ B}\delta
(\sqrt{\bA})^B_{ \ C}(
(\sqrt{\bA})^{-1})^C_{ \ D}=\delta
(\bA)^A_{ \ B} ((\sqrt{\bA})^{-1})^B_{
\ D} \ .
\end{equation}
Taking the trace of given equation we
immediately obtain (\ref{defdeltatrA}).
Note also that due to the matrix nature
of objects $\bA$ and $\bB$ the
following relation is not valid
\begin{equation}\label{sqrtAB}
\sqrt{\bA \bB}=\sqrt{\bA}\sqrt{\bB} \
\end{equation}
unless $\bA$ and $\bB$ commute. On the
other hand  since obviously $\bA$ and
$\bA^{-1}$ commute the equation
(\ref{sqrtAB}) gives
\begin{equation}
\sqrt{\bA}\sqrt{\bA^{-1}}=\mathbf{I} \
\end{equation}
which implies following important
relation
\begin{equation}
\left(\sqrt{\bA}\right)^{-1}=\sqrt{\bA^{-1}}
\ .
\end{equation}}
\begin{equation}\label{defdeltatrA}
\delta \sqrt{\bA^A_{ \ A}}= c_1\delta
\bA^A_A+2c_1 \delta \bA^A_{ \ B}\bA^B_{
\ A}+3c_3 \delta \bA^A_{ \ B}\bA^B_{ \
C}\bA^C_{ \ A}+\dots=\frac{1}{2} \delta
\bA^A_{ \ B}
\left((\sqrt{\bA})^{-1}\right)^B_{ \ A}
\ .
\end{equation}
% since
%\begin{equation}
%\sqrt{\bA\bB}\sqrt{\bA\bB}=\bA\bB
%\neq \sqrt{\bA}\sqrt{\bB}\sqrt{\bA}\sqrt{\bB}
%\end{equation}
%but of course this relation is valid under trace.
For that reason we restrict ourselves
to the Hamiltonian analysis of the
non-linear massive gravity action with
the potential term
\begin{equation}
\mU=4\tr \sqrt{\bA}\ .
\end{equation}
In fact, the Hamiltonian analysis of
the gauge fixed form of this non-linear
massive gravity was firstly  performed
in \cite{Kluson:2011qe} and then in
\cite{Hassan:2011ea,Golovnev:2011nz}.
It was shown in these two
papers that this theory possesses interesting
properties that allow to
eliminate one additional scalar mode.
Then our goal is to analyze given theory
written in the manifest diffeomorphism
invariant form and try to identify
possible additional constraints
that could eliminate one scalar mode.
Explicitly,  we would like to perform
the Hamiltonian analysis of the
following action
\begin{equation}\label{Smassive1}
S=M_p^2\int d^3\bx dt  N
\sqrt{g}\left[K_{ij}\mG^{ijkl}K_{kl}+{}^{(3)}R
- m^2 \tr \sqrt{\bA}\right] \ .
\end{equation}

%%%%%%%%%%%%%%%%%%%%%%%%%%%%%%%%%%%%%%%5
\section{Hamiltonian
Analysis}\label{third}
 In this section
we perform the Hamiltonian analysis of
the action (\ref{Smassive1}). For the
General Relativity part of the action
the procedure is standard. Explicitly,
the momenta conjugate to $N,N_i$ are
the primary constraints of the theory
\begin{equation}
\pi_N(\bx)\approx 0 \ , \quad
\pi^i(\bx)\approx 0 \
\end{equation}
while the Hamiltonian takes the form
\begin{eqnarray}\label{HamGR}
H^{GR}&=&\int d^3\bx
(N\mH^{GR}_T+N^i\mH^{GR}_i) \ ,
\nonumber \\
\mH_T&=&\frac{1}{\sqrt{g}M_p^2}
\pi^{ij}\mG_{ijkl}\pi^{kl}-M_p^2
\sqrt{g} {}^{(3)}R \ , \nonumber \\
\mH_i&=& -2g_{ik}
\nabla_l\pi^{kl} \ , \nonumber \\
\end{eqnarray}
where $\pi^{ij}$ are momenta conjugate
to $g_{ij}$ with following non-zero
Poisson brackets
\begin{equation}
\pb{g_{ij}(\bx),\pi^{kl}(\by)}=
\frac{1}{2}\left(\delta_i^k\delta_j^l+\delta_i^l
\delta_j^k\right)\delta(\bx-\by) \ .
\end{equation}
Finally note that $\pi_N,\pi^i$ have
following Poisson brackets with $N,N_i$
\begin{equation}
\pb{N(\bx),\pi_N(\by)}=\delta(\bx-\by)
\ , \quad \pb{N_i(\bx),\pi^j(\by)}=
\delta_i^j\delta(\bx-\by) \ .
\end{equation}
Now we proceed to the Hamiltonian
analysis of the scalar field part of
the action. Note that in $3+1$
formalism the matrix $\bA^A_{ \ B}$
takes the form
\begin{equation}
\bA^A_{\
B}=-\nabla_n\phi^A\nabla_n\phi_B+
g^{ij}\partial_i\phi^A\partial_j\phi_B
\equiv K^A_{ \ B}+V^A_{ \ B} \ ,
\end{equation}
where
\begin{eqnarray}
K^A_{ \
B}&=&-\nabla_n\phi^A\nabla_n\phi_B \ ,
\quad K_{AB}=\eta_{AC}K^C_{ \ B}=K_{BA}
\ , \nonumber \\
V^A_{ \
B}&=&g^{ij}\partial_i\phi^A\partial_j\phi_B
\ , \quad V^{AB}=V^A_{ \ C}\eta^{CB}=
V^{BA} \ . \nonumber \\
\end{eqnarray}
Then the  conjugate momenta $p_A$ are
equal to
\begin{eqnarray}\label{pAext}
p_A&=&-\frac{M_p^2
m^2}{2}\sqrt{g}\frac{\delta \bA^C_{ \
D}}{\delta
\partial_t\phi^A}(\bA^{-1/2})^D_{ \
C}=\nonumber
\\
&=&\frac{M_p^2m^2}{2}\sqrt{g}(\nabla_n\phi_C
(\bA^{-1/2})^C_{ \ A}+
\eta_{AC}(\bA^{-1/2})^C_{ \
B}\nabla_n\phi^B) \ , \quad
\bA^{-1/2}=(\sqrt{\bA})^{-1} \ .
\nonumber
\\
\end{eqnarray}
Note that using the symmetry of
$\bA_{AB}=\bA_{BA}$ we can write
(\ref{pAext})  in simpler form
\begin{equation}
p_A=M_p^2
m^2\sqrt{g}(\bA^{-1/2})_{AB}\nabla_n\phi^B
\ .
\end{equation}
Using this expression we derive
following relation
%and we have
%\begin{eqnarray}
%\frac{1}{g}p_Ap^A=(M^{-1})^{CD}(g^{ij}\partial_i\phi_D\partial_j\phi_C-M_{DC})=
%(M^{-1})^{CD}g^{ij}\partial_i\phi_D\partial_j\phi_C-\tr
%\mathbf{I}
%\nonumber \\
%\end{eqnarray
\begin{eqnarray}
\frac{1}{gM_p^4 m^4}p_A p_B&=&
(\bA^{-1/2})_{AC}(\nabla_n\phi^C\nabla_n\phi^D)
(\bA^{-1/2})_{DB}= \nonumber \\
&=& (\bA^{-1/2})_{AC}( V^{CD}-\bA^{CD})
(\bA^{-1/2})_{DB} \nonumber \\
%&=& (\sqrt{M}^{-1})_{AC} V^{CD}
%(\sqrt{M}^{-1})_{DB}-\sqrt{M}_A^{ \ D}
%(\sqrt{M})^{-1}_{DB}= \nonumber \\
% \Rightarrow  \nonumber \\
%& &(\frac{1}{g}p_A
%p_B+\eta_{AB})V^{BE}=(\sqrt{M}^{-1})_{AC}V^{CD}
%(\sqrt{M}^{-1})_{DB}V^{BE} \ .  \Rightarrow  \nonumber \\
\end{eqnarray}
which implies
\begin{equation}\label{PiAV}
\Pi_{AB}=(\bA^{-1/2})_{AC}V^{CD}(\bA^{-1/2})_{DB}
\ ,
\end{equation}
where we introduced the matrix
$\Pi_{AB}$ defined as
\begin{equation}
\Pi_{AB}=\frac{1}{gm^4 M_p^4}p_A
p_B+\eta_{AB} \ .
\end{equation}
Note that when we multiply (\ref{PiAV})
by $V$ from the right we obtain (we use
matrix notation)
\begin{equation}
\Pi V=(\bA^{-1/2}V)(\bA^{-1/2}V)
\end{equation}
which  implies
\begin{equation}\label{bAV1}
\bA^{-1/2}V= \sqrt{\Pi V} \ .
\end{equation}
This relation will be important below.
The important point of our analysis is
that $V^{AB}$ has the rank $3$ as was
firstly explicitly stressed in
\cite{Hassan:2012qv}. In fact, if we
introduce the $4\times 3$ matrix $W^A_{
\ i}=
\partial_i\phi^A$ and its transpose
matrix $(W^T)^i_{ \
A}=\partial_i\phi^A$ which is $3\times
4$ matrix we can write
\begin{equation}
V^{AB}=W^A_{ \ i}g^{ij}(W^T)_j^{ \ B} \
.
\end{equation}
Then since $W^A_{ \ i},g^{ij}$ have the
rank $3$ we obtain that $V^{AB}$ has
the rank $3$ as well. As a result $\det
V=0$. In other words $V$ is not
invertible matrix. This fact has an
fundamental consequence for the
Hamiltonian structure of given theory.
% Note
%that in case of the existence of
%$V^{-1}$ the equation (\ref{bAV1})
%gives
%\begin{equation}\label{bA1piV}
%\bA^{-1/2}=\sqrt{\Pi V} V^{-1} \ .
%\end{equation}
On the other hand we can multiply
(\ref{PiAV}) by $V$ from the left so
that
\begin{eqnarray}
%\Pi_{AB}= (\sqrt{M}^{-1})_{AC}V^{CD}
%(\sqrt{M}^{-1})_{DB}
% \Rightarrow  \nonumber \\
V\Pi=(V\bA^{-1/2})(V\bA^{-1/2})
\nonumber \\
\end{eqnarray}
that now implies
\begin{equation}
\sqrt{V\Pi}= V\bA^{-1/2} \ .
\end{equation}
With the help of these results it is
easy to determine corresponding
Hamiltonian
\begin{eqnarray}\label{HamSC}
\mH^{sc}&=&\partial_t \phi^A
p_A-\mL_{sc}
%N\nabla_n\phi^A
%p_A-\mL_{matt}+N^i\partial_i\phi^A p_A=
%\nonumber \\
%=-M_p^2m^2\sqrt{g}K^{AB}(\bA^{-1/2})_{BA}-
%\mL_{matt}+N^i\mH_i^{sc}=\nonumber \\
%=-m^2 M_p^2\sqrt{g}(\bA^{AB}-V^{AB})
%(\bA^{-1/2})_{BA}+M_p^2m^2\sqrt{g}N(\bA^{1/2})^A_A
%+N^i\mH_i^{sc}=
%\nonumber \\
=M_p^2m^2\sqrt{g}NV^{AB}(\bA^{-1/2})_{BA}+N^ip_A\partial_i\phi^A=
%\sqrt{g}N(\bA^{-1/2})^{AB}V_{BA}+N^i\mH_i^{sc}=
\nonumber \\
&=&NM_p^2m^2\sqrt{g}\sqrt{\Pi_{AB}V^{BA}}+N^i
p_A\partial_i\phi^A\equiv
N\mH_T^{sc}+N^i\mH_i^{sc} \   \nonumber \\
\end{eqnarray}
using (\ref{bAV1}) and using an obvious
relation $\tr \sqrt{V\Pi}= \tr
\sqrt{\Pi V}$.  With the help of these
results we find the final form of the
Hamiltonian formulation of the action
(\ref{Smassive1})
\begin{equation}
H=\int d^3\bx (N\mH_T+N^i\mH_i) \ ,
\end{equation}
where
\begin{equation}
\mH_T=\mH_T^{GR}+\mH_T^{sc} \ , \quad
\mH_i=\mH_i^{GR}+\mH_i^{sc} \ ,
\end{equation}
where the explicit forms of these terms
is given in (\ref{HamGR}) and in
(\ref{HamSC}). Then note that the
requirement of the preservation of the
primary constraints $\pi_N\approx 0 \ ,
\pi^i\approx 0$ implies an existence of
the secondary constraints
\begin{equation}\label{seccon}
\mH_T(\bx)\approx 0 \ , \quad
\mH_i(\bx)\approx 0 \ .
\end{equation}
The crucial question is the existence
of additional constraints in the
theory. The existence of these
constrains would be important  for the
elimination of an additional scalar
mode. We should expect that this
 constraint is the primary
constraint between momenta and
coordinates of the scalar fields and
the question is whether we can find
such a relation. For example, we can
try to calculate $\det \Pi$. Using
\begin{eqnarray}
& &\det
\left(\eta_{AB}+\frac{1}{gM_p^4m^4}p_A
p_B\right)=
% \det \eta_{AC}\det
%(\delta^C_B+\frac{1}{g}\eta^{CD}p_Ap_B)=
\det \eta_{AC} \det \left(\delta^C_{ \
B}+\frac{1}{gM_p^4m^4}p^Cp_B\right)=
\nonumber \\
& &=\det \eta
 \exp \tr \ln
\left(\delta^A_{ \
B}+\frac{1}{gM_p^4m^4}p^A p_B\right)=
%=\det \exp (\frac{1}{g}
%p_Ap^A-\frac{1}{2g^2}(p_Ap^A)^2+\dots)=
\det \eta \exp \ln\left
(1+\frac{1}{gM^4_pm^4} p_A p^A\right)=
\nonumber \\
&=&
% =\det \eta (1+\frac{1}{g}p_A p^A)=
-\left(1+\frac{1}{gM_p^4m^4}p_A
p^A\right)
 \nonumber \\
\end{eqnarray}
and hence from (\ref{PiAV}) we obtain
\begin{eqnarray}\label{detPi}
%\det
%(\frac{1}{g}p_Ap_B+\delta_{AB})=\frac{\det
%V}{\det M} \Rightarrow  \nonumber \\
1+\frac{1}{gM_p^4m^4} p_A
p^A=-\frac{\det V}{\det \bA} \ .
\nonumber \\
\end{eqnarray}
We see that the upper equation implies
the primary constraint on condition
when $\det V=0$.
% In fact, this
%condition is automatically obeyed in
%two dimensions since
%\begin{equation}
%\det V=V^{00}V^{11}-V^{01}V^{01}=
%g^{11}(\partial_1 \phi^0)^2
%g^{11}(\partial_1 \phi^1)^2-
%\left(g^{11}\partial_1
%\phi^0\partial_1\phi^1\right)^2= 0 \ .
%\end{equation}
Then (\ref{detPi}) implies the primary
constraint
\begin{equation}
\mC: \frac{1}{gM^4_pm^4}p_Ap^A+1=0 \ .
\end{equation}
The fact that $\det V^{AB}=0$ is
however crucial for the calculation of
the algebra of the constraints.
Unfortunately we are not able to
perform this calculation in case of the
four dimensional case due to the
absence of the inverse matrix $V^{-1}$.
We demonstrate the complexity of such a
calculation in the appendix when we
abandon the condition $\det V^{AB}=0$.
However we are able to perform this
calculation in case of two dimensional
massive gravity which we perform in the
next section.

\section{Two
Dimensional Massive Gravity}
\label{fifth} Two dimensional massive
gravity is exceptional also from the
fact that the gravity is trivial so
that the dynamical content is hidden in
the scalar degrees of freedom only. In
order to determine the physical number
of degrees of freedom we have to find
an appropriate structure of
constraints.
 To proceed we
have to determine  the form of the
Hamiltonian constraint (\ref{HamSC}).
An explicit calculation gives
\begin{eqnarray}
\mH_T&=&M_p^2m^2\sqrt{g}\tr\sqrt{\Pi
V}=
%\sum_n \tr c_n (\Pi V)^n =\nonumber \\
\sqrt{\frac{1}{\omega}(p_A\partial
\phi^A)^2+M_p^4m^4\partial
\phi_A\partial\phi^A} \ , \nonumber \\
%\mH_S=p_A\partial\phi^A-2\omega\nabla \pi^\omega \  \nonumber \\
\end{eqnarray}
%\sum_n c_n (A+C)^n= \sqrt{A+C}
%=\nonumber \\
%=\sqrt{(p_0\partial
%\phi^0+p_1\partial\phi^1)^2-g^{11}(\partial
%\phi^0)^2+g^{11} (\partial \phi^1)^2}
% \nonumber \\
%\end{eqnarray}
%using
%\begin{eqnarray}
%\Pi V= \left(\begin{array}{cc}
%\Pi_{00}V^{00}+\Pi_{01}V^{01} &
%\Pi_{00}V^{01}+\Pi_{01}V^{11} \\
%\Pi_{01}V^{00}+\Pi_{11}V^{01} &
%\Pi_{10}V^{01}+\Pi_{11}V^{11}
%\\ \end{array}\right)\equiv
%\left(\begin{array}{cc}
%A & B \\
%D & C \\ \end{array}\right) \ , \nonumber \\
%\Pi_{00}V^{00}= (p_0\partial
%\phi^0)^2-V^{00} \ ,
%\Pi_{11}V^{11}=(p_1\partial\phi^1)^2+V^{11}
%\ ,
%\Pi_{01}V^{10}=(p_0\partial\phi^0)(p_1\partial\phi^1)
%, \nonumber \\
%\tr \Pi V=A+C=(p_0\partial
%\phi^0+p_1\partial\phi^1)^2+V^{00}+V^{11}
%\nonumber \\
%(\Pi V)^2= \left(\begin{array}{cc}
%A^2+BD & AB+BC \\
%DA+CD & DB+C^2 \\ \end{array}\right) \
%, \tr (\Pi V)^2=
%(A+C)^2+2(BD-AC)=(A+C)^2 \nonumber \\
%\tr (\Pi V)^3= (A+C)^3+3A(BD-AC)+
%3C(BD-AC)=(A+C)^3 \ , \nonumber \\
%\end{eqnarray}
%since $BD-AC=0$.
%
%As a result in two dimensional case the
%Hamiltonian constraint has the form
%\begin{equation}
%\mH_T=-\sqrt{g}\sqrt{\frac{1}{g}(p_0\partial
%\phi^0+p_1\partial\phi^1)^2+g^{11}(\partial
%\phi^0)^2+g^{11} (\partial \phi^1)^2}
%\end{equation}
where we introduced  following notation
\begin{equation}
g_{11}=\omega \ , \quad  \det g=\omega
\ , \quad  g^{11}=\frac{1}{\omega} \ ,
\quad
\partial_1=\partial \ .
\end{equation}
As a result the extended  Hamiltonian
takes the form
\begin{eqnarray}
H&=& \int dx (N\mH_T+N^1\mH_S+
u^\omega\pi^\omega+u^N\pi_N+ u_1\pi^1+
\Gamma \mC) \ ,  \nonumber \\
\mH_T&=&\sqrt{\frac{1}{\omega}(p_A\partial
\phi^A)^2+M_p^4m^4\partial
\phi_A\partial\phi^A}\equiv \sqrt{\bA}
\ , \quad
\mH_S=p_A\partial\phi^A-2\omega\nabla \pi^\omega \ , \nonumber \\
\nonumber \\
\end{eqnarray}
where we included the primary
constraint $\pi^\omega\approx 0$ into
the definition of the spatial
diffeomorphism constraint. Note the
this theory possesses following four
primary constraints
\begin{equation}
\pi_N\approx 0 \ , \quad \pi^1\approx 0
\ , \quad  \pi^\omega\approx 0 \  ,
\quad \mC=\frac{1}{\omega
M_p^4m^4}p_Ap^A+1=0 \ .
\end{equation}
First of all the requirement of the
preservation of the primary constraints
$\pi_N,\pi^1$ implies following
secondary ones
\begin{equation}
\mH_T\approx 0 \ , \quad  \mH_S\approx
0 \ .
\end{equation}
We see that $\mH_T\approx
0,\mH_S\approx 0$ are the secondary
constraints. Then we have to check the
consistency of all primary and
secondary constraints during the time
evolution of the system. The time
evolution of the constraints
$\pi_N,\pi^1$ does not generate new
conditions while
 the requirement of
the preservation of the constraint
$\pi^\omega $ implies
\begin{eqnarray}\label{partpiomega}
\partial_t\pi^\omega&=&\pb{\pi^\omega,H}=
\frac{N(p_A\partial\phi^A)^2}{\omega^2\sqrt{\bA}}
-\Gamma \triangle=\nonumber \\
&=&\frac{N(\mH_S+2\omega\nabla
\pi^\omega)^2}{\omega^2\sqrt{\bA}}
-\Gamma \triangle\approx -\Gamma
\triangle=0 \ ,  \nonumber \\
\end{eqnarray}
where we used following Poisson bracket
\begin{equation}\label{deftriangle}
\pb{\mC(\bx),\pi^\omega(\by)}
=\frac{1}{\omega^2}p_Ap^A(\bx)\delta(\bx-\by)\equiv
\triangle(\bx) \delta(\bx-\by) \
\end{equation}
and we used the fact that we have to
determine the evolution on the
constraint surface. On the other hand
the time evolution of the constraint
$\mC$ is equal to
\begin{equation}\label{timemC}
\partial_t\mC=\pb{\mC,H}\approx
\pb{\mC,\bT_T(N)}+\int dx u^\omega
\pb{\mC,\pi^\omega(x)}= \nonumber \\
\pb{\mC,\bT_T(N)}+u^\omega \triangle=0
\ ,
\end{equation}
where we used the fact that
\begin{equation}
\pb{\bT_S(N^1),\mC}= -N^1\partial \mC
\end{equation}
which vanishes on the constraint
surface and where $\pb{\mC,\bT_T(N)}$
is equal to
\begin{eqnarray}
\pb{\bT_T(N),\mC}= 2\partial
\left(\frac{N (p_A\partial\phi^A)p_B}
{\omega\sqrt{\bA}}\right)
\frac{p^B}{\omega}+ 2\partial\left(N
\frac{\partial
\phi^A}{\sqrt{\bA}}\right)\frac{p_A}{\omega}
\ . \nonumber \\
%\pb{\mC(\bx),\pi^\omega(\by)}
%=\frac{1}{\omega^2}p_Ap^A(\bx)\delta(\bx-\by)\equiv
%\triangle(\bx) \delta(\bx-\by) \ .
%\nonumber \\
\end{eqnarray}
Note that this Poisson bracket vanishes
on the constraint surface $\mH_S\approx
0 \ , \pi^\omega\approx 0$ as well. As
a result we find that
(\ref{partpiomega}) and (\ref{timemC})
determine that Lagrange multipliers
$u^\omega,\Gamma$ are equal to zero.

Finally we have to determine the
algebra of constraints $\mH_T,\mH_S$.
As usual we calculate the Poisson
brackets of their smeared forms and we
find
\begin{eqnarray}
\pb{\bT_T(M),\bT_T(N)}= \frac{1}{4}\int
dx dy \frac{N(x)}{\sqrt{\bA}(x)}
\pb{\bA(x),\bA(y)}\frac{N(y)}{\sqrt{\bA(y)}}
\ . \nonumber \\
\end{eqnarray}
To proceed we calculate following
Poisson bracket
\begin{eqnarray}
& &\pb{\bA(x),\bA(y)}=\nonumber \\
&=& -4\frac{1}{\omega(x)}
(p_A\partial\phi^A)(x)\left[
\partial\phi^A(x)p_A(y)
\partial_y\delta(x-y)-\right.\nonumber \\
& &- \left.
p_A(x)\partial\phi^A(y)\partial_x
\delta(x-y)\right]\frac{1}{\omega(y)}
(p_A\partial\phi^A)(y)+
\nonumber \\
&+&4\frac{M_p^4m^4}{\omega(x)}
(p_A\partial\phi^A)(x)\left[
(\partial\phi^A(x)\partial\phi_A(y)
\partial_y\delta(x-y)
-\right.
\nonumber \\
& & -\left.
(\partial\phi^A(y)\partial\phi_A(x)
\partial_x\delta(x-y)\right]\frac{1}{\omega(y)}
(p_A\partial\phi^A)(y) \ .
\nonumber \\
\end{eqnarray}
In the calculation of the Poisson
bracket of the smeared Hamiltonian
constraints we use the fact the Poisson
bracket $\pb{\bA(x),\bA(y)}$ contains
derivative of the delta functions that
give non-zero contribution when they
act on $N$ and $M$ respectively. As a
result we obtain
\begin{eqnarray}
\pb{\bT_T(M),\bT_T(N)}&=& \int dx
(M\partial N-N\partial M)
\frac{1}{\bA\omega}
\times \nonumber \\
&\times & [ (p_A\partial\phi^A)^3
+M_p^4m^4(p_A\partial\phi^A)(
(\partial\phi^A\partial\phi_A)]=
%=\int d\bx (M\partial N-N\partial M)
%\frac{1}{\omega \bA} [\bA
%(p_A\partial\phi^A)]
%% \int d\bx
%(N\partial M-M\partial N)\mH_S^{sc}=
%=\nonumber \\
\nonumber \\
&=&\bT_S\left(\frac{1}{\omega}(N\partial
M-M\partial N)\right)+2\int dx
(N\partial M-M\partial N)\omega \nabla
\pi^\omega \ .
\nonumber \\
\end{eqnarray}
The algebra of constraints $\bT_S$
takes the standard form
\begin{equation}
\pb{\bT_S(M^1),\bT_S(N^1)}=
\bT_S(M^1\partial N^1-N^1\partial M^1)
\ .
\end{equation}
Finally we determine the Poisson
bracket
\begin{eqnarray}
\pb{\bT_S(M^1),\bT_T(N)}&=& \int dx
\pb{\bT_S(M^1),\mH_T(x)}N=\nonumber
\\
&=&-\int dx(M^1\partial\mH_T+
\mH_T\partial X^1)=
 \bT_T(M^1\partial N)  \nonumber \\
\end{eqnarray}
using
\begin{equation}
\pb{\bT_S(M^1),\mH_T(x)}=-M^1\partial\mH_T-
\mH_T\partial M^1
 \ .
\end{equation}

 Finally we have to
analyze the stability of the
constraints $\mH_T$ and $\mH_S$. In
fact, the constraint $\bT_S(M^1)$ is
preserved due to the fact that the
Poisson brackets between all
constraints vanish on the constraint
surface. On the other hand the time
evolution of the constraint is given by
the equation
\begin{eqnarray}
\partial_t \bT_T(M)=
\pb{\bT_T(M),H}\approx \int dx
(u^\omega\pb{\bT_T(M),\pi^\omega}
+\Gamma \pb{\bT_T(M),\mC})=0
 \nonumber
\\
% \int dx
%(u^\omega\pb{\bT_T(X),\pi^\omega}\approx
%0 \nonumber \\
\end{eqnarray}
using the fact that $u^\omega$ and
$\Gamma$ vanish.

At this stage we have finished the
analysis of the time development of the
constraints with following result. We
have four first class constraints
\begin{equation}
\pi_N\approx 0 \ , \quad  \pi^1\approx
0 , \quad  \mH_T \approx 0 \ , \quad
\mH_1\approx 0
\end{equation}
and the second class constraints
\begin{equation}
\pi^\omega \approx 0 \ , \quad
\mC\approx 0 \ .
\end{equation}
Using the second class constraints we
can eliminate $\pi^\omega $ and
$\omega$ as functions of $p_A$. Then
the gauge fixing of the four first
class constrains completely eliminate
the physical degrees of freedom
$N,\pi_N,N^1,\pi_1$ and $p_A,\phi^A$.

\section{Conclusion}\label{sixth}
In this section we outline our results and suggest the possible
extension of this work. We performed the Hamiltonian analysis of
some particular model of non-linear massive gravity action written
in the St\"{u}ckelberg picture \cite{deRham:2011rn}.
 We found
corresponding Hamiltonian. Then following \cite{Hassan:2012qv} where
it was explicitly shown an existence of the primary and
corresponding secondary constraint in the version of the non-linear
massive gravity action presented in \cite{Golovnev:2011nz} we find
corresponding primary constraint of the theory. Unfortunately due to
the fact that this constraint is a consequence of the singularity of
the matrix $V^{AB}$ we were not able to determine algebra of all
constraints and identify secondary constraints for the case of four
dimensional non-linear massive gravity action. On the other hand we
were able to complete the Hamiltonian analysis of two dimensional
non-linear massive gravity where we showed that the algebra of the
Hamiltonian and spatial diffeomorphism constraints
 is in agreement with the basic principles of geometrodynamics
\cite{Kuchar:1974es,Isham:1984sb,Isham:1984rz}. We also identified
an additional constraints and we show that there are no physical
degrees of freedom with agreement with the analysis presented in
\cite{deRham:2011rn,Kluson:2011aq}.

It is very unhappy that we were not able to finish the Hamiltonian
analysis of  four dimensional non-linear massive gravity theory due
to the singular nature of the matrix $V^{AB}$ especially in the
light of the very nice proof of the existence of the primary and the
secondary constraints that was performed in \cite{Hassan:2012qv} in
the case of the  version of the non-linear massive gravity action
presented in \cite{Golovnev:2011nz}. To finish such an analysis is
very desirable and it is the main goal of our future work. It would
be also very nice to perform the Hamiltonian analysis of the general
form of the non-linear massive gravity action with
 the St\"{u}ckelberg fields and we hope to return to this problem in
 future.

%\footnote{We would like to stress that
%the results derived in this paper has
%closed overlap with  the Hamiltonian
%analysis of the Higgs mechanism of
%gravity
%\cite{'tHooft:2007bf,Chamseddine:2011mu,Alberte:2010it,Chamseddine:2010ub,Iglesias:2011it}
%that was performed in
%\cite{Kluson:2010qf}.}.
 %Unfortunately
%we were not able to find  additional
%primary constraints whose presence
%could lead to the constraint structure
%that would be able to eliminate one
%additional scalar mode. We mean however
%that this  is very puzzling fact that
%while the gauge fixed form of the
%theory has the correct number of the
%physical degrees of freedom its
%St\"{u}ckelberg formulation does not
%seem to be able to eliminate
%non-physical mode. Unfortunately we are
%not able to give an explanation of this
%issue.  Of course there is a
%possibility that we miss some primary
%constraints between canonical variables
%or it is possible that  for the
%potential presented in
%\cite{deRham:2010kj} such a constraint
%exists. However the analysis of this
%more general theory   is rather
%intricate and we leave it for the
%future. It is also crucial to find an
%explanation why our results do not
%agree with analysis performed in
%\cite{Hassan:2011ea,Golovnev:2011nz}.
\vskip 4mm

\begin{appendix}

\section{Toy Model: Algebra of
Constraints in Case of $\det V^{AB}\neq
0$ }\label{appendix} In this appendix
we perform the calculation of the
algebra of constraints in case when we
do not impose the condition $\det
V^{AB}=0$ even if these calculations
are not directly related to the case of
the non-linear massive gravity action
studied in the main body of this paper.
The goal of this appendix is to
demonstrate the complexity of given
analysis. To begin with note that the
fact that $\det V^{AB}\neq 0$ implies
 the existence inverse
matrix $V^{-1}$. In this case we find
an important relation
\begin{equation}\label{comPiV}
V^{-1}\sqrt{V\Pi}=\bA^{-1/2}= \sqrt{\Pi
V}V^{-1} \
\end{equation}
which  will be useful when we calculate
the algebra of constraints.

 Let us
consider the  smeared form of the
constraints (\ref{seccon})
\begin{equation}\label{smearedconst}
\bT_T(N)=\int d^3\bx N\mH_T \ , \quad
\bT_S(N^i)=\int d^3\bx N^i\mH_i \ .
\end{equation}
The goal of this section is to
determine Poisson brackets among these
constraints.
%It is convenient to consider their
%smeared form
%\begin{equation}
%\bT_T(N)=\int d^d\bx N\mH_T \ , \quad
%\bT_S(N^i)=\int d^d\bx N^i\mH_i
%\end{equation}
Note that in case of the General
Relativity part of the constraints we
have following Poisson brackets
\begin{eqnarray}\label{PBGRCON}
\pb{\mH_T^{GR}(\bx),\mH_T^{GR}(\by)}&=&
-\left[\mH^i_{GR}(\bx)\frac{\partial}{\partial
x^i}
\delta(\bx-\by)-\mH^i_{GR}(\by)\frac{\partial}{\partial
y^i} \delta(\bx-\by)\right] \ , \nonumber \\
\pb{\mH_T^{GR}(\bx),\mH_i^{GR}(\by)}&=&
\mH^{GR}_T(\by)\frac{\partial}{\partial
x^i}\delta(\bx-\by) \ , \nonumber \\
\pb{\mH^{GR}_i(\bx),\mH_j^{GR}(\by)}&=&
\left[\mH_j^{GR}(\bx)\frac{\partial}{\partial
x^i} \delta(\bx-\by)-\mH_i(\by)
\frac{\partial}{\partial y^j}
\delta(\bx-\by)\right] \ . \nonumber \\
\end{eqnarray}
The calculation of the Poisson brackets
that contains scalar phase space
degrees of freedom is more involved.
However it is easy to find the Poisson
bracket between generators of spatial
diffeomorphisms
\begin{equation}
\pb{\mH^{sc}_i(\bx),\mH_j^{sc}(\by)}=
\left[\mH_j^{sc}(\bx)\frac{\partial}{\partial
x^i} \delta(\bx-\by)-\mH^{sc}_i(\by)
\frac{\partial}{\partial y^j}
\delta(\bx-\by)\right] \
\end{equation}
that together with the  Poisson bracket
on the third line in  (\ref{PBGRCON})
implies following form of Poisson
bracket between smeared form of the
diffeomorphism constraints
\begin{equation}\label{pbbTSS}
\pb{\bT_S(N^i),\bT_S(M^j)}=
\bT_S(N^j\partial_j M^i-M^j\partial_j
N^i) \ .
\end{equation}
 In case of
Hamiltonian constraint the situation is
not so easy. Explicitly, from the
definition of the Poisson bracket we
find
%\begin{eqnarray}
%\pb{\mH_T^{sc}(\bx),\mH_T^{sc}(\by)}=
%\pb{\sqrt{g}\sqrt{p_Ap_B+\delta_{AB}}\sqrt{V}^{BA}(\bx),
%\sqrt{g}\sqrt{p_Cp_D+\delta_{CD}}\sqrt{V}^{DC}(\by)}=
%\nonumber \\
%=-\int d^D\bz \left(
%\sqrt{g}\frac{\delta
%(\sqrt{g^{-1}p_Ap_B+\delta_{AB}}(\bx))}
%{\delta p_X(\bz)}\sqrt{V}^{BA}(\bx)
%\frac{\delta \sqrt{V}^{DC}(\by)}{\delta
%\phi^X(\bz)}\sqrt{g}\sqrt{g^{-1}p_Cp_D+\delta_{CD}}(\by)-
%\right. \nonumber \\
%\left. - \sqrt{g}\frac{\delta
%(\sqrt{g^{-1}
%p_Cp_D+\delta_{CD}}(\by))}
% {\delta p_X(\bz)}\sqrt{V}^{DC}(\by)
%\frac{\delta \sqrt{V}^{BA}(\bx)}{\delta
%\phi^X(\bz)}\sqrt{g}
%\sqrt{g^{-1}p_Ap_B+\delta_{AB}}(\bx)
%\right)
%\nonumber \\
%\end{eqnarray}
%It is convenient to calculate the
%smeared form of these constraints
\begin{eqnarray}\label{bTNM}
& &
\pb{\bT^{sc}_T(N),\bT^{sc}_T(M)}=\int
d^3\bx d^3\by N(\bx)
\pb{\mH_T^{sc}(\bx),\mH_T^{sc}(\by)}M(\by)=
\nonumber \\
&=&-M^4_p m^4\int d^3\bx d^3\by
 d^3\bz
 N(\bx)M(\by) \left(
\sqrt{g}(\bx) \frac{\delta (\sqrt{\Pi
V})_{A}^{ \ A}(\bx)} {\delta p_X(\bz)}
\frac{\delta (\sqrt{\Pi V})_C^{\
C}(\by)}{\delta
\phi^X(\bz)}\sqrt{g}(\bx) -
\right. \nonumber \\
& &\left. - \sqrt{g}(\by)\frac{\delta
(\sqrt{\Pi V})_{C}^{ \ C}(\by)}
 {\delta p_X(\bz)}
\frac{\delta (\sqrt{\Pi V})_A^{\
A}(\bx)}{\delta
\phi^X(\bz)}\sqrt{g}(\bx) \right) \ .
\nonumber \\
\end{eqnarray}
To proceed further we use the relations
\begin{eqnarray}\label{helprelation}
\frac{\delta (\sqrt{\Pi V})_{A}^{ \
A}(\bx)}{\delta p_X(\bz)}&=&
\frac{1}{2} \frac{\delta (\Pi V)_A^{ \
B}(\bx)}{\delta p_X(\bx)}((\Pi
V)^{-1/2})_B^{ \ A}=
\nonumber \\
%\frac{\delta (\sqrt{\Pi})_{AB}(\bx)}
%{\delta p_X(\bz)}=
% \frac{1}{2gM_p^4m^4}
%\frac{\delta(
% p_Ap_C+\delta_{AC})(\bx)}
%{\delta
%p_X(\bz)}(\sqrt{\Pi^C_B}^{-1}(\bx)=
%\nonumber \\
&=&\frac{1}{2gM_p^4m^4} (\delta^X_A
p_C+p_A\delta^X_C)V^{CB} \left((\Pi
V)^{-1/2}\right)_B^{\
A}(\bx)\delta(\bx-\bz) \ ,
\nonumber \\
\frac{\delta (\sqrt{V \Pi})^C_{ \
C}(\by)} {\delta \phi^X(\bz)}&=&
\frac{1}{2}\frac{\delta (V\Pi)^C_{ \
D}(\by)} {\delta \phi^X(\bz)}
((V\Pi)^{-1/2})^D_{ \ C}(\by)=
\nonumber \\
&=&\frac{1}{2} [g^{ij}
\partial_{y^i}\delta(\bz-\by)
\delta_X^C\partial_{y^j}\phi^E(\by)+
g^{ij}\partial_{y^i}\phi^C(\by)\partial_{y^j}
\delta_X^E\delta(\bz-\by)]\Pi_{ED}\left((V\Pi)^{-1/2}\right)^D_{
\ C}(\by) \ .
\nonumber \\
\end{eqnarray}
Then with the help of
(\ref{helprelation}) we can determine
the Poisson bracket (\ref{bTNM}). First
of all these Poisson brackets contain
the derivative of the delta functions.
We perform integration by parts. Then
the non-zero contribution arises from
the derivative of the smeared functions
$N$ and $M$. As a result we obtain
\begin{eqnarray}
\pb{\bT^{sc}_T(N),\bT^{sc}_T(M)}
%=\frac{1}{4}\int d^D\bx
%(N\partial_iM-M\partial_i
%N)g^{ij} \times \nonumber \\
%\times (\delta^X_A p_F+p_A\delta^X_F)
%(\sqrt{p^Fp_B+\eta^{FD}\delta_{DB}})^{-1}
%\sqrt{V}^{BA} \times \nonumber \\
%\times  (
%\delta_X^D\partial_{y^j}\phi^E+
%\partial_{y^j}\phi^D
%\delta_X^E)(\sqrt{V}^{-1})^C_E
%\sqrt{p_Cp_D+\delta_{CD}}=
%\nonumber \\
&=&\frac{1}{4}\int d^3\bx
(N\partial_iM-M\partial_i
N)g^{ij} \times \nonumber \\
&\times & \left( p_A(\sqrt{(V\Pi
)^{-1}}V)^{AB}+
p_A(V\sqrt{(\Pi V)^{-1}})^{AB}\right) \times \nonumber \\
&\times &  (
(\Pi\sqrt{(V\Pi)^{-1}})_{BC}\partial_j\phi^C+
(\sqrt{(\Pi V)^{-1}}\Pi)_{BC}\partial_j
\phi^C)=\nonumber \\
%=\frac{1}{4}\int d^D\bx
%(N\partial_iM-M\partial_i
%N)g^{ij} \times \nonumber \\
%\times ( p_F
%(\sqrt{\Pi^{-1}}\sqrt{V})^{FX}+
%(p_A\sqrt{V}\sqrt{\Pi^{-1}})^{AX}) \times \nonumber \\
%\times  (
%(\sqrt{\Pi}\sqrt{V^{-1}})_{XE}\partial_j\phi^E+
%(\sqrt{V^{-1}}\sqrt{\Pi})_{XD}\partial_j
%\phi^D)=\nonumber \\
&=&\int d^3\bx
(N\partial_iM-M\partial_i N)g^{ij}
 p_A
(V\sqrt{(\Pi V)^{-1}})^{AB}(\sqrt{(\Pi
V)^{-1}}\Pi)_{BC}\partial_j
\phi^C)= \nonumber \\
&=&\int d^3\bx
(N\partial_iM-M\partial_i N)g^{ij}p_A
(V(\Pi V)^{-1}\Pi)^A_{ \
B}\partial_j\phi^B=\nonumber
\\
&&=\bT^{sc}_S((N\partial_jM-M\partial_j
N)
g^{ji}) \ , \nonumber \\
\end{eqnarray}
where
 we used the symmetry of $\Pi$
and $V$ and  the relations
\begin{equation}
\sqrt{(V\Pi)^{-1}}V=V\sqrt{(\Pi
V)^{-1}}, \quad \Pi\sqrt{(V\Pi)^{-1}}=
\sqrt{(\Pi V)^{-1}}\Pi
\end{equation}
 that follow from
(\ref{comPiV}).
 If we combine
this result with the Poisson brackets
between smeared form of the General
Relativity Hamiltonian constraints we
find the final result
\begin{equation}\label{pbbTTf}
\pb{\bT_T(N),\bT_T(M)}= \bT_S(
(N\partial_jM-M\partial_j N) g^{ji}) \
.
\end{equation}
%Then we have
%\begin{eqnarray}
%\int d^3\bz \frac{\delta
%(\sqrt{p_Ap_B+\delta_{AB}}(\bx))}
%{\delta p_X(\bz)}\sqrt{V}^{BA}(\bx)
%\frac{\delta \sqrt{V}^{DC}(\by)}{\delta
%\phi^X(\bz)}\sqrt{p_Cp_D+\delta_{CD}}(\by)=
%\nonumber \\
%\frac{1}{4}\int d^3\bz (\delta^X_A
%p_C(\bx)+p_A\delta^X_C)
%(\sqrt{p^Cp_B+\eta^{CD}\delta_{DB}})^{-1}(\bx)\delta(\bx-\bz)
%\sqrt{V}^{BA}(\bx) \times
%\nonumber \\
%(g^{ij}
%\partial_{y^i}\delta(\bz-\by)
%\delta_X^D\partial_{y^j}\phi^E(\by)+
%g^{ij}\partial_{y^i}\phi^D(\by)\partial_{y^j}
%\delta_X^E\delta(\bz-\by))(\sqrt{V}^{-1})^C_E(\by)
%\sqrt{p_Cp_D+\delta_{CD}}(\bx)
%\nonumber \\
%\end{eqnarray}
Finally we calculate the Poisson
bracket between $\bT_S(N^i)$ and
$\bT^{sc}_T(N)$. Using
\begin{eqnarray}
\pb{\bT_S(N^i),p_A(\bx)}&=&-N^i\partial_ip_A(\bx)-
\partial_i N^ip_A(\bx) \ , \nonumber \\
\pb{\bT_S(N^i),\phi^A(\bx)}&=&-N^i\partial_i\phi^A(\bx)
\ ,
\nonumber \\
\pb{\bT_S(N^i),\sqrt{g}(\bx)}&=&-N^i\partial_i
\sqrt{g}(\bx)-\partial_i
N^i\sqrt{g}(\bx) \ ,  \nonumber \\
\pb{\bT_S(N^i),g^{ij}(\bx)}&=&
-N^k\partial_k g^{ij}(\bx)+\partial_k
N^i g^{kj}(\bx)+g^{ik}\partial_k
N^j(\bx) \ ,
\nonumber \\
\pb{\bT_S(N^i),\Pi_{AB}(\bx) }&=&
-N^i\partial_i \Pi_{AB}(\bx) \ ,
\nonumber \\
%\pb{\bT_S(N^i),\partial_i\phi^A(\bx)}&=&
%-\partial_i N^k\partial_k\phi^A(\bx)-
%N^k\partial_k (\partial_i\phi^A)(\bx) \
%,
%\nonumber \\
\pb{\bT_S(N^i),V^{AB}(\bx)}&=&-N^k\partial_k
V^{AB}(\bx) \ . \nonumber \\
\end{eqnarray}
With the help of these results it is
easy to find
\begin{eqnarray}
\pb{\bT_S(N^i),\mH_T^{sc}(\bx)}
%=
%(\sqrt{g}\partial_i N^i+\partial_i
%\sqrt{g}N^i) \tr \sqrt{\Pi}\sqrt{V}-
%\nonumber \\
%-\sqrt{g}N^k\partial_k\left(
%\frac{1}{g}p_A p_B\right)\frac{\delta
%\sqrt{\Pi}\sqrt{V}}{\delta \Pi_{BA}}
%-\sqrt{g}\sqrt{\Pi}N^k\partial_k V^{AB}
%\frac{\delta \sqrt{V}}{\delta V^{BA}}=
%\nonumber \\
=-\partial_k N^k\mH^{sc}_T(\bx)-
N^k\partial_k \mH^{sc}_T(\bx) \ .
\nonumber
\\
\end{eqnarray}
Collecting this result with the Poisson
bracket between diffeomorphism
constraint and Hamiltonian constraint
of General Relativity we obtain
\begin{equation}\label{bTTSf}
\pb{\bT_S(N^i),\bT_T(N)}=
\bT_T(\partial_k N N^k) \ .
\end{equation}
To conclude, we found that the theory
with $\det V\neq 0$  possesses four
constraints $\mH_T(\bx)\approx 0 \ ,
\mH_i(\bx)\approx 0$ which are the
first class constraints as  follows
from the Poisson brackets
(\ref{pbbTSS}),(\ref{pbbTTf}) and
(\ref{bTTSf}). We also showed that the
crucial presumption for these
calculations was the regularity of the
matrix $V^{AB}$ which of course is not
the case of four dimensional non-linear
massive gravity.

\end{appendix}
 \noindent {\bf
Acknowledgement:} I would like to thank
F. Hassan for very useful discussion.
This work was supported by the Czech
Ministry of Education under Contract
No. MSM 0021622409. \vskip 5mm

\end{document}